\documentclass[aps,prb,twocolumn,showpacs]{revtex4}

\usepackage{amssymb}
\usepackage{graphicx}
\usepackage{dcolumn}
\usepackage{bm}

\def\be{\begin{equation}}
\def\ee{\end{equation}}
\def\bea{\begin{eqnarray}}
\def\eea{\end{eqnarray}}

\begin{document}
\title{Suppressing phonon transport in nanowires: a simple model for phonon-surface roughness interaction}
\author{K. A. Muttalib and S. Abhinav}
\affiliation
{Department of Physics, University of Florida, Gainesville, FL 32611-8440, USA}
%\today
\begin{abstract}

Suppressing phonon propagation in nanowires is an essential goal towards achieving efficient thermoelectric devices. Recent experiments have shown unambiguously that surface roughness is a key factor that can reduce the thermal conductivity well below the Casimir limit in thin crystalline silicon nanowires. We use insights gained from the experimental studies to construct a simple analytically tractable model of the phonon-surface roughness interaction that provides a better theoretical understanding of the effects of surface roughness on the thermal conductivity, which could potentially help in designing better thermoelectric devices. 

\end{abstract}


\maketitle
 
\section{Introduction}

To convert waste heat to electricity or to use electricity for refrigeration, an efficient thermoelectric material needs to have a large electrical conductivity $\sigma$ and at the same time a poor thermal conductivity $\kappa$ \cite{slack,takabatake}. Typical bulk materials with large $\sigma$ are Fermi liquids, and they turn out to be inherently inefficient  \cite{reviews,snyder}  because the ratio of the two conductivities at a given temperature, $\kappa/\sigma T$ where $T$ is the temperature, is a fixed number independent of the material properties  (the Wiedemann-Franz law) \cite{AM}. On the other hand it has been widely recognized that these two conductivities can be controlled independently in nano-engineered materials, boosting the efficiency \cite{hicks,majumdar,datta1,dresselhaus,reddy}. A typical nanodevice consists of two large leads at different temperatures
connected by a quantum dot \cite{sothman,nakanishi,ws,jordan}, a molecule \cite{koch,wacker,murphy,aharony,finch}, or a 
wire \cite{hicks3,pichard,kim}, and one can, in principle, have finite power output at high efficiency \cite{hmn,whitney,linke} in such systems. In particular it has been shown that thin disordered Si nanowires can be used to achieve both high efficiency and large power output \cite{mh} provided one operates in the non-linear regime and takes advantage of an interplay between the microscopic parameters of the material and the thermodynamic parameters of the leads as well as the connected load \cite{hmn}. Since Si nanowires are reliably reproducible, and the device is  not very sensitive to fine tuning of microscopic parameters, this looks highly promising. However, one factor that has not been considered in such systems in a systematic and quantitative way is the effects of phonons. Indeed, a large contribution to $\kappa$ from phonons can seriously reduce the thermoelectric efficiency of any device.

While heat conduction in molecular junctions \cite{mingo,galperin} as well as disordered waveguides \cite{gil} and wires \cite{mingoNL,mingo2003,wal,nika} has been considered theoretically in some detail, there has been a series of recent experimental studies on Si nanowires \cite{boukai,hochbaum,li,lim,heron,blanc1,blanc2} that show the importance of surface roughness in wires with diameters $d< 150$ nm. Li \textit{et al} \cite{li} measures the thermal conductivity $\kappa$ as a function of temperature $T$  from $25 - 325$ K for a series of `smooth' Si nanowires (grown by vapor-liquid-solid process) of diameters $115$, $56$, $37$ and $22$ nm. The  dependence of $\kappa$ on the diameters of the wires (with the exception of the $22$ nm wire) can be understood in terms of Boltzmann transport of phonons through a tube with specular as well as diffuse boundary scattering \cite{mingoBT}.
However as shown by Hochbaum \textit{et al} \cite{hochbaum},  if the wire is prepared in a different way (electroless etching nanowires) such that the surface is `more rough', even the maximum diffusive surface scattering model can not explain the phonon thermal conductance which can be almost an order of magnitude smaller near room temperature. In fact, the thermal conductivity in such  wires can reach the amorphous limit when the diameter  $d \sim 50$ nm, even though the wire is far from being amorphous.  Apparently this surprisingly small $\kappa$ can be explained within a Born approximation for phonon scattering where the surface roughness changes the phonon dispersion relation \cite{martin}. However, it has been argued that such an approximation should break down at wavelengths comparable to the size of the scatterers. Indeed, an atomic level investigation \cite{carrete} concludes that Born approximation overestimates thermal resistance by an order of magnitude, and so can not explain the experiments of Hochbaum \textit{et al}. On the other hand Monte Carlo simulations \cite{moore,lacroix} show possible phonon mean free paths below the Casimir limit (of the order of the diameter), and Molecular Dynamics simulations \cite{donadio,he,zushi,liu} for very thin wires ($d\le 15$ nm) show significant reduction in the thermal conductivity in the presence of an oxide layer, an amorphous surface or a periodic ripple roughness. While these atomistic numerical simulations illustrate the importance of surface roughness on the thermal conductivity, a simple analytically tractable model that captures the essentials of the phonon-surface roughness interaction as suggested by the experiments has not been considered yet.

Lim \textit{et al} \cite{lim} have done a systematic characterization of the surface roughness to understand the difference in the two sets of wires studied in Ref. [\onlinecite{hochbaum}], and concludes that a frequency dependent phonon scattering from rough surfaces is important.    
In this work we use these insights to construct and develop a simple model of phonon-surface roughness interaction and use standard analytical techniques to calculate its effects on phonon scattering rate. The calculations are done at the simplest level of lowest order perturbation theory at zero temperature, but the robust features of the model naturally lead to a frequency dependent phonon scattering whose strength depends on the peak value of the roughness power spectrum. We find that while a constant form of the roughness power spectrum washes out the signature of the frequency dependent scattering in the transmission function, using the experimentally determined form of the power spectrum \cite{lim} leads to specific features that can be tested experimentally. In addition, we identify an important parameter of the model with a particular combination of the parameters related to the rms surface roughness profile and the roughness correlation length. The thermal conductivity at a fixed temperature decreases significantly as a function of this disorder parameter.  

\section{Model for phonon-surface roughness interaction}

In order to study the effects of surface scattering in the context of electron transport in thin films, Te$\check{s}$anovi\'{c} et al introduced an exact mapping that allowed a reformulation of the problem in terms of a smooth surface with an effective interaction Hamiltonian that includes channel-mixing pseudo-potential terms \cite{tesanovich}:
\bea
H_{eff} &=& \sum_{k,n}(\xi_q+E_n)c^{\dag}_{k,n}c_{k,n} \cr
&+& \sum_{k,q}\sum_{m,n}\lambda(q) V_{mn} c^{\dag}_{k+q,n}c_{k,m}.
\eea
Here  $c^{\dag}$, $c$  are the electron creation and annihilation operators, respectively, and the matrix $V_{mn}$ contains channel mixing terms $(E_n-E_m)[z\partial/\partial z + (\partial/\partial z)z]_{mn}$ where $E_n$ are the eigenvalues of the film with smooth surface parallel to the x-y plane.

Adapting the same mapping to phonon transport in a thin wire, it is clear that the channel mixing term proportional to the coordinate $z$ will correspond to a \textit{localized} phonon operator, which couples to the bulk propagating phonons. The position of the localized phonons must be randomly distributed if the surface roughness is random. Since it is not possible to obtain either the surface roughness spectrum or the coupling of the localized phonons from the mapping without detailed inputs about the microscopic properties of the surface, we propose a simple model where inputs from experiments can be used to obtain a qualitative understanding of the effects of phonon-surface roughness interactions.
In particular, we assume that a phonon Hamiltonian with extended imperfections on the surface will have a pseudo-potential $\phi(x-x_l)$  where the site $x_l$ related to the localized phonon are randomly distributed. The Hamiltonian for the propagating phonons interacting with the surface roughness can then be written as 
\bea
H_{int} = \sum_l\int\;dx A(x)A(x)\phi(x-x_l)
\eea
where the phonon operator $A(x)=b(x)+b^{\dag}(x)$, $b$ and $b^{\dag}$ being the usual destruction and creation operators for the propagating phonons. We expect $\phi$ to be proportional to $\xi=a+a^{\dag}$ where $a$ and $a^{\dag}$ are the destruction and creation operators for localized phonons arising from surface disorder.  Moreover, a correlation of the type $\langle \phi \phi\rangle_{random}$ would include a measure of the root mean square fluctuations of the thickness of the wire. We will implement these ideas as we develop the model. For simplicity, we will use $\hbar=1$. 

Within the standard perturbation theory \cite{AGD} for the time-ordered Green function 
\be
D(x,x')\equiv  -i\langle T[A(x)A(x')]\rangle, 
\ee
the lowest non-zero contribution 
\bea
&&\langle \delta^{(2)}D(x,x') \rangle_{random} 
= \frac{1}{2}\sum_{l,m}\int dx_1dx_2 \cr
&& \langle T[A(x)A(x')A(x_1)A(x_1) A(x_2)A(x_2)]\rangle F(x_1,x_2) 
\eea
involves a thermal as well as disorder average  
\bea
&& F(x_1,x_2)\equiv \left\langle\langle T\phi (x_1-x_l)\phi (x_2-x_m)\rangle\right\rangle_{random} \cr
&&= \sum_{q_1,q_2}e^{iq_1x_1} e^{iq_2x_2}\langle T\phi_{q_1}(t_1)\phi_{q_2}(t_2)\rangle\cr
&&\times \left \langle \sum_{l,m}e^{-iq_1x_l} e^{-iq_2x_m} \right\rangle_{random}.
\label{PhiPhi}
\eea
(A first order contribution leads to an average over a single phonon operator $\phi_q$, which vanishes.) We now use the standard techniques used for impurity averaging of electrons in a disordered media \cite{doniach}, namely  
\bea
\left\langle \sum_{l,m}e^{-iq_1x_l} e^{-iq_2x_m}\right\rangle_{random} &=& N_{imp}\delta_{q_1+q_2,0} \cr
&+& N_{imp}^2\delta_{q_1,0}\delta_{q_2,0}
\eea
and use the idea mentioned above that $\phi_q$ is proportional to the localized phonon operator, 
\be
\phi_q(t)=u_q\xi_q(t). 
\ee
Then 
\bea
F(x_1,x_2)= i N_{imp}\sum_{q_1}e^{iq_1(x_1-x_2)}u_{q_1}u_{-q_1}d_{q_1}(t_1-t_2) 
\label{Phi}
\eea
where we defined the impurity averaged time-ordered localized phonon Green function
\be
d_q(t_1-t_2) \equiv -i\langle T[\xi_q(t_1)\xi_{q}(t_2)]\rangle 
\ee
and kept only the term linear in the impurity number density $N_{imp}$. 
Wick's theorem  then allows us to obtain the exchange self-energy \cite{AGD}
\bea
\Sigma_p(\omega)= N_{imp}\sum_{q,\nu}  u_{q}u_{-q}d_{q}(\nu) D^0_{p-q}(\omega-\nu)
\label{sigp}
\eea
where $D^0_p(\omega)$ is the non-interacting time-ordered propagating Green function. For explicit calculations later, we note that the non-interacting non-equilibrium lesser and greater  Green functions have the form \cite{rammer}
\bea  
D^{0,\lessgtr}_p(\omega)=-i2\pi\left[(1+ N_p)\delta(\omega\pm\omega_p)+N_p\delta(\omega\mp\omega_p) \right]
\eea
where $N_p$ is the number of phonons at momentum $p$ and we assume an acoustic dispersion relation $\omega_p=vp$, $v$ being the sound velocity. On the other hand if the surface disorder is strong, we expect it to lead to strongly localized phonons, but not necessarily strong coupling to the propagating phonons. Thus 
 the impurity averaged localized non-equilibrium lesser or greater phonon Green functions can be taken to have the form
\bea
d^{\lessgtr}_q(\nu)=i2\Gamma_q\left[\frac{1+N_q}{(\nu\pm\Omega_q)^2+\Gamma_q^2}+ \frac{N_q}{(\nu\mp\Omega_q)^2+\Gamma_q^2}\right]
\eea
where $\Omega_q$ is the frequency and $\Gamma_q$ is the width of the localized phonon. Since the localized phonons arise from surface disorder, we expect a broad distribution of $\Omega_q$ including modes inside the propagating phonon band.

To connect with experiments, we need an appropriate model for $u_qu_{-q}$ in (\ref{sigp}). For a thin wire, a confining potential in the transverse direction implies that $u_q$ should be proportional to $1/Md^2$ where $M$ is the effective mass of an atom at the surface and $d$ is the wire thickness \cite{tesanovich}. It should also be proportional to a measure of surface disorder related to the fluctuations of the wire thickness. We write
\be
u_qu_{-q}=\frac{\lambda^2}{M^2d^4} W(q)
\ee
where $\lambda$ is a dimensionless strength of the coupling between the localized and the propagating phonons assumed to be small, and $W(q)$ is proportional to the roughness power spectrum $S(q)$ which is a key measure of the roughness of the wire \cite{lim,sadhu}.

% RELAXATION TIME:
%====================

\section{The relaxation time}

Use of second order perturbation theory should be a good start for qualitative predictions in our model since we assume weak coupling of the propagating phonons to strongly localized phonons.
The most significant effect of the phonon-surface roughness interaction is to give rise to a single particle relaxation time $\tau_i(\omega)$ for the propagating phonons, given by the imaginary part of the retarded self energy. 
The non-equilibrium Green function formulation \cite{rammer} allows us to write the imaginary part simply as
$2 {\rm Im}\;\Sigma_p^R(\omega)=i[\Sigma_p^<(\omega)-\Sigma_p^>(\omega)]=-1/\tau_i(\omega)$. 
The lesser and greater self energies for the three phonon processes (propagating phonon emitting and absorbing a localized phonon) have the form
\bea
i\Sigma_p^{\lessgtr}(\omega) &=& 2 \frac{N_{imp}\lambda^2}{m^2d^4}\sum_qW(q) J^{\lessgtr}_{p,q}(\omega)\cr
J^{\lessgtr}_{p,q}(\omega) & \equiv & \int_{-\infty}^{\infty}d\nu\; d^{\lessgtr}_q(\nu)D_{p-q}^{0,\lessgtr}(\omega-\nu)
\label{siglessgtr}
\eea
where the factor $2$ takes into account two possible diagrams that turn out to be equivalent\cite{mingo}. The calculation is particularly simple at zero temperature:
\bea
J^{<}_{p,q}(\omega) - J^{>}_{p,q}(\omega) &=& \frac{4\pi\Gamma_q}{(\omega+\Omega_q+\omega_{p-q})^2+\Gamma_q^2}\cr
& - &\frac{4\pi\Gamma_q}{(\omega-\Omega_q-\omega_{p-q})^2+\Gamma_q^2}.
\label{sigma1}
\eea

The summation over the momenta $q$ in (\ref{siglessgtr}) will depend on how we model the momentum dependence of various parameters.  The experiments suggest that the crucial momentum dependence arises from the power law form of the roughness power spectrum, which we must keep. We simplify our model by ignoring all other momentum dependence, $\Omega_q\approx \Omega$ and $\Gamma_q\approx \Gamma$. For the relaxation time calculation, we will consider the effect of a single localized phonon in the propagating phonon band, which can be easily extended to arbitrary number of localized phonons. (Effects of many such phonons with a distribution of $\Omega$ will be explored while considering the transmission function in section IV.)  
According to Lim et al \cite{lim}, $S(q)$ is best described by a power law, $S(q)\propto (q_0/q)^n$ with $n$ between $2$ and $3$, in the range of wave vector $q_1=0.01\; {\rm to}\; q_2=1\; \rm{(nm)^{-1}}$ while it is approximately constant for $q \le q_1$. Then the relaxation time $\tau_i$ due to the phonon-surface roughness interaction can be written as
\bea
&&\frac{1}{2\tau_i(\omega)} = \frac{4\pi N_{imp}\lambda^2 \rho_0}{M^2d^4}\int_{0}^{\infty} d\varepsilon W(\varepsilon) R(\varepsilon) \cr
&&R(\varepsilon) \equiv   \frac{\Gamma}{(\omega-\Omega-\varepsilon)^2+\Gamma^2}-\frac{\Gamma}{(\omega+\Omega+\varepsilon)^2+\Gamma^2}
\label{sigma3}
\eea
where we have replaced the sum over the momenta by an integral over the (positive) phonon energies $\varepsilon$ with a constant density of states $\rho_0$, and used the model for the roughness power spectrum
\bea
W(\varepsilon) = W_0 \left\{ \begin{array}{rcl} (\varepsilon_0/\varepsilon_1)^n;  &\mbox{for}& \varepsilon \le \varepsilon_1, \\
(\varepsilon_0/\varepsilon)^n;  & \mbox{for} & \varepsilon_1 <\varepsilon < \varepsilon_0  .
\end{array}\right.
\label{W}
\eea
For the sake of definiteness, we will assume $\varepsilon_0 \gg \Omega \gg \Gamma \gg \varepsilon_1$, without loss of generality. 

\begin{figure}
\includegraphics[angle=0,width=0.42\textwidth]{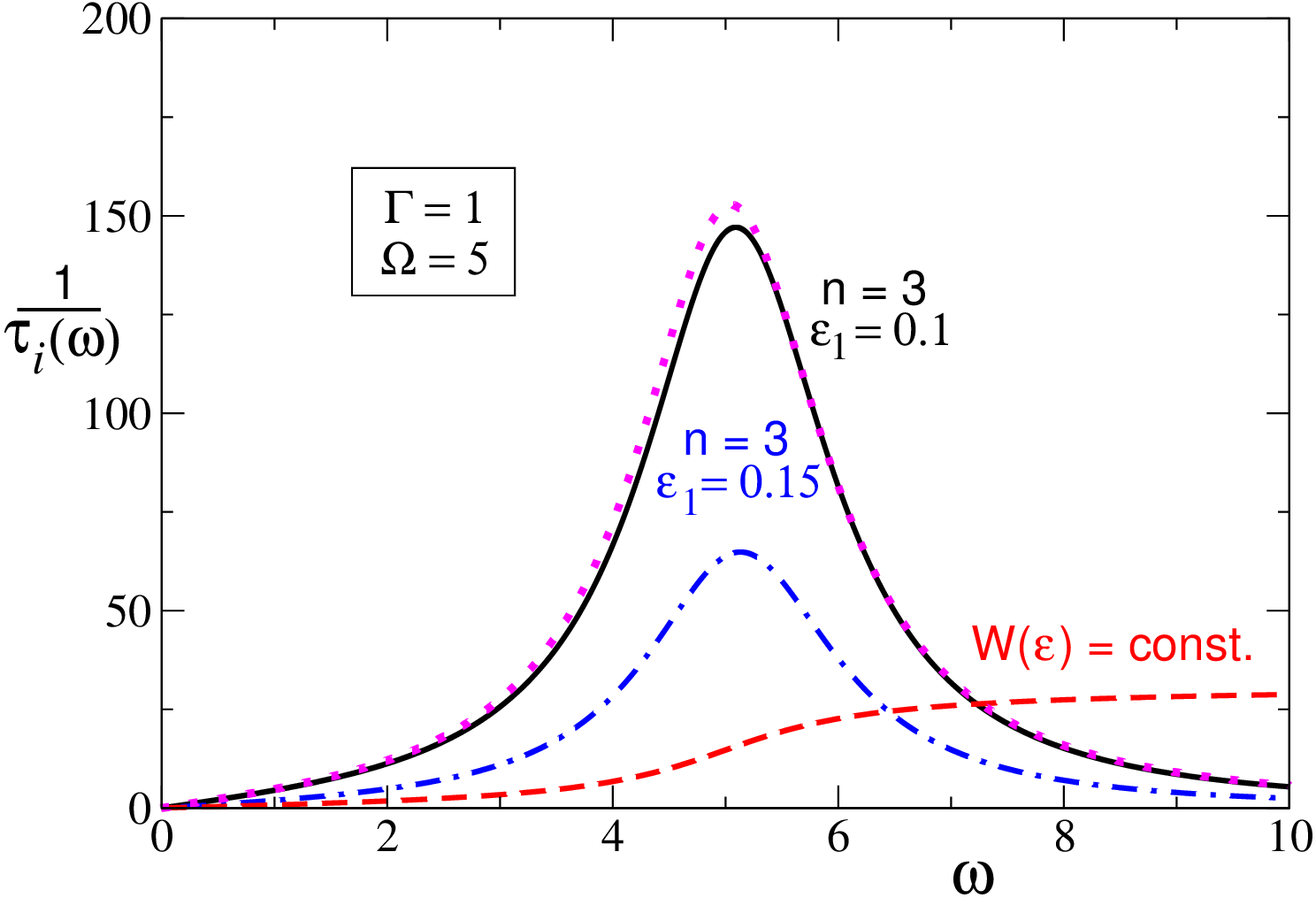}
\caption{ Single particle relaxation rate $1/\tau_i(\omega)$, in arbitrary units,  for two different choices of the roughness power spectrum parameter $\varepsilon_1$ defined in (\ref{W}) compared with a Lorentzian (dotted (magenta) line) and a constant $W(\varepsilon)$ (dashed (red) line). All energies are in units of $\Gamma_0$.}
\label{fig1}
\end{figure}

We first note that the contribution from $W(\varepsilon)$ is larger for larger exponent $n$, and the experiment suggests $2 < n < 3$ for many different types of disorder. For simplicity, we will use $n=3$ for all our illustrations. The large contribution from small $\epsilon$ in $W(\varepsilon)$ is largely offset from $R(\varepsilon)$ by the small factor $\omega\Gamma/\Omega^3$ in regime where $\omega\ll \Omega$, and by the small factor $\Omega\Gamma/\omega^3$ in the regime where $\omega \gg \Omega$. In contrast, this contribution is enhanced by the large factor $1/\Gamma$ in the regime where $\omega\approx \Omega$. This gives rise to an energy dependent scattering which was first established experimentally in  Ref.~[\onlinecite{lim}] and plays a crucial role in our current model. The physical picture is simple; propagating phonons with frequency that matches the frequency of the localized phonons are scattered the most, the effectiveness of the scattering depending on how sharply defined the localized phonons are.  Figure \ref{fig1} shows the phonon-surface roughness scattering relaxation rates for two different choices of the roughness power spectrum parameter $\varepsilon_1$ in units of the localized phonon parameter $\Gamma=\Gamma_0$. Note that only a comparison among the two is intended; no attempt has been made to fix the magnitude which depends on the unknown prefactor in (\ref{sigma3}). 
Clearly, the surface scattering is highly sensitive to the roughness parameter. The result is qualitatively different if we choose $W(\varepsilon)$ to be a constant, as shown in Figure \ref{fig1}, where the relaxation rate becomes flat at large $\omega$. In this case for illustrative purposes we have chosen $W(\varepsilon)=W_0(\Gamma_0^3/\varepsilon_0^2\varepsilon_1)$ which would be the value at $\varepsilon=0$ for $n=1$, but the general frequency dependence does not depend on this particular choice.  

An alternative model for the power spectrum is a Lorentzian \cite{martin,goodnick}, of the form
\be
W(\varepsilon) = \Delta^2 \frac{\varepsilon_c}{\varepsilon^2+\varepsilon^2_c}
\ee
where $\varepsilon_c$ is inversely proportional to the disorder correlation length $L_c$, and $\Delta$ is proportional to the rms surface profile. We expect $\Delta$ to be proportional to the ratio of the mean surface roughness height $h$ and the diameter $d$ of the wire, $\Delta\propto h/d$.
In order to make a comparison with the power law model, we choose the value  $W(\varepsilon_1)$ to be the same for both models. We expect $\varepsilon_c \ll \varepsilon_1$; for the purpose of an explicit illustration we choose $\varepsilon_c$ to be smaller than $\varepsilon_1$ by a factor 5 and adjust $W_0$ by the same factor in order to obtain the same value of the the power spectrum at $\varepsilon=0$. The result is shown by the dotted (magenta) curve in Figure \ref{fig1}. Thus the Lorentzian and the power-law models give very similar results with appropriately chosen parameters. 

The Lorentzian model allows us to identify the ratio $(\varepsilon_0/\varepsilon_1)$ with 
$L_c$ and $\Delta$ to match the value $W(\varepsilon=\varepsilon_1)$, namely 
\be
\left(\frac{\varepsilon_1}{ \varepsilon_0}\right)^{n-2}\propto\frac{L_c}{\Delta^2},
\ee
where we have used $\varepsilon_c\propto 1/L_c$.  It is known that as $\Delta$ increases and/or $L_c$ decreases, the thermal conductivity decreases. The above relation is then consistent with the fact that a decrease in $\varepsilon_1$ increases the scattering rate.

% TRANSMISSION FUNCTION:
%========================

\section{Transmission function}

A potential signature of the validity of the model is the presence of localized phonons within the phonon band, which should be observable in the frequency dependent transmission function $\mathcal{T}(\omega)$ given by 
\be
\mathcal{T}(\omega)=\sum_p\Lambda_L D^R_p(\omega)\Lambda_R D^A_p (\omega).
\ee
Here $\Lambda_{L,R}$ are the couplings to the left and right leads which, for simplicity, we take to be independent of $\omega$, and $D^{R,A}$ are the `dressed' retarded and advanced propagating phonon Green functions that include the scattering off the localized phonons as a self energy contribution. This leads to 
\bea
\mathcal{T}(\omega)=\sum_p \frac{2\omega_p\Lambda_L}{[\omega+\frac{i}{2\tau(\omega)}]^2-\omega_p^2}\frac{2\omega_p\Lambda_R}{[\omega-\frac{i}{2\tau(\omega)}]^2-\omega_p^2}
\eea
where $\tau(\omega)$ is the total relaxation time due to all scattering mechanisms present  in a nanowire. This includes not only the $\tau_i$ due to the phonon-surface roughness interactions calculated above, but also scatterings from bulk impurities, anharmonic terms, umklapp or boundary scattering etc. In particular, the umklapp scatterings are essential in order to explain the saturation of the thermal conductivity as a function of temperature $T$ at room temperature. For our purposes, in order to illustrate the effects of the phonon-surface roughness interaction, we will ignore all other scattering mechanisms except the boundary scattering that contributes a small constant term $\tau_b$. The reason for keeping the boundary scattering term is that for a smooth thin wire without any surface disorder, the boundary scattering contributes a small but constant relaxation rate, typically proportional to the inverse power of the diameter of the wire \cite{mingo2003}. This is in addition to any scattering due to surface disorder and leads to a finite transmission at $\omega=0$. Ignoring the umklapp scattering in particular will limit us to the low-$T$ regime only.
Nevertheless, keeping only $\tau_i$ and $\tau_b$ would allow  us to obtain important low-temperature properties of the thermal conductivity considered in the next section. We expect the above two scattering processes to be independent such that the two inverse scattering times can be added together according to Matthiessen's rule.  As before, replacing the sum over momentum by the corresponding energy integral, we obtain  
\bea
\mathcal{T}(\omega)=2\pi\rho_0\Lambda_L\Lambda_R\tau(\omega)\propto \frac{\tau_b\tau_i(\omega)}{\tau_b+\tau_i(\omega)}.
\eea

As seen in the previous section, a localized phonon with a given frequency $\Omega$  in the propagating band will contribute to an anomalously large scattering of propagating phonons around $\Omega$. For more than one localized phonon with frequencies $\Omega_i$, $i=1,2 \cdots $, the transmission function should develop minima around each of the $\Omega_i$, which are expected to be randomly distributed due to disorder.
However, depending on the width of the localized phonons, it may or may not be possible to identify the individual phonons from the transmission function. As a simple but illustrative example, we  consider the effect of five localized phonons within the range $ \Gamma_0 < \Omega <20$, arbitrarily chosen at $\Omega=3,5,8,13$ and $14$ in units of $\Gamma_0$. We will choose $\tau_b$ such that $\mathcal{T}(\omega)=1$ at $\omega=0$. 

\begin{figure}
\includegraphics[angle=0,width=0.42\textwidth]{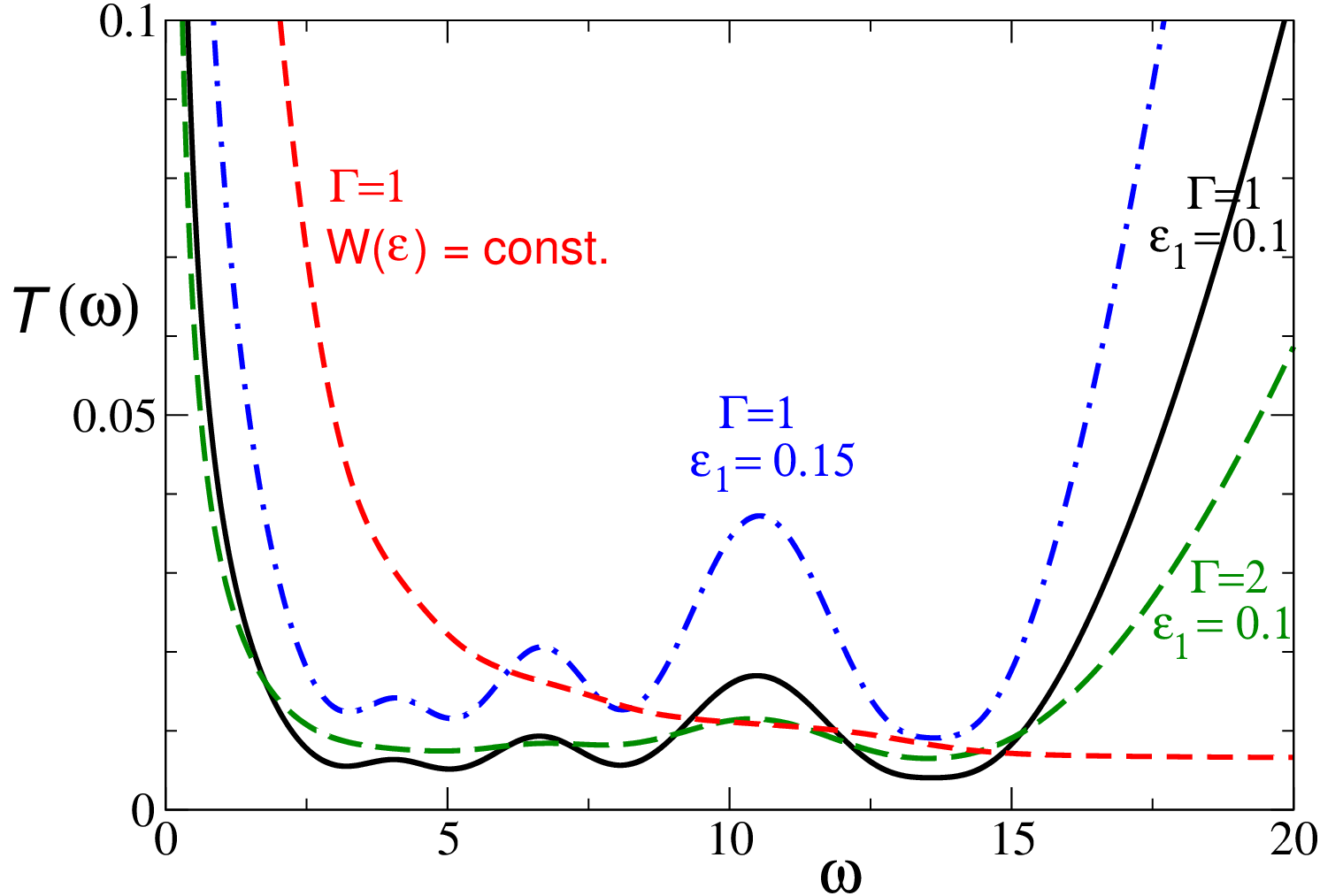}
\caption{Transmission function $\mathcal{T}(\omega)$, in arbitrary units, for $n=3$ with different combinations of the parameters $\Gamma$ and $\varepsilon_1$, compared with a constant $W(\varepsilon)$ (dashed (red) line). All energies are in units of $\Gamma_0$.}
\label{fig2}
\end{figure}

Figure \ref{fig2} shows the transmission function for three different combinations of $\varepsilon_1$ and $\Gamma$ (assumed to be the same for all of the five modes) for $n=3$, compared with a constant $W(\varepsilon)$.
While most phonon modes are identifiable for small $\Gamma=1$, increasing $\Gamma$ tends to wash out the individual phonon dips. On the other hand, a constant power spectrum function wipes out all structures in the transmission function, as shown by the (red) dashed line in Figure \ref{fig2}. Thus
careful experiments on the frequency dependence of the transmission function and detailed analysis based on a realistic distribution of the localized phonon parameters should allow one to determine the characteristics of any localized phonons present and how they change with surface disorder, as well as the role and importance of the power spectrum function.

% THERMAL CONDUCTIVITY:
%========================

\section{Thermal conductivity}

\begin{figure}
\includegraphics[angle=0,width=0.42\textwidth]{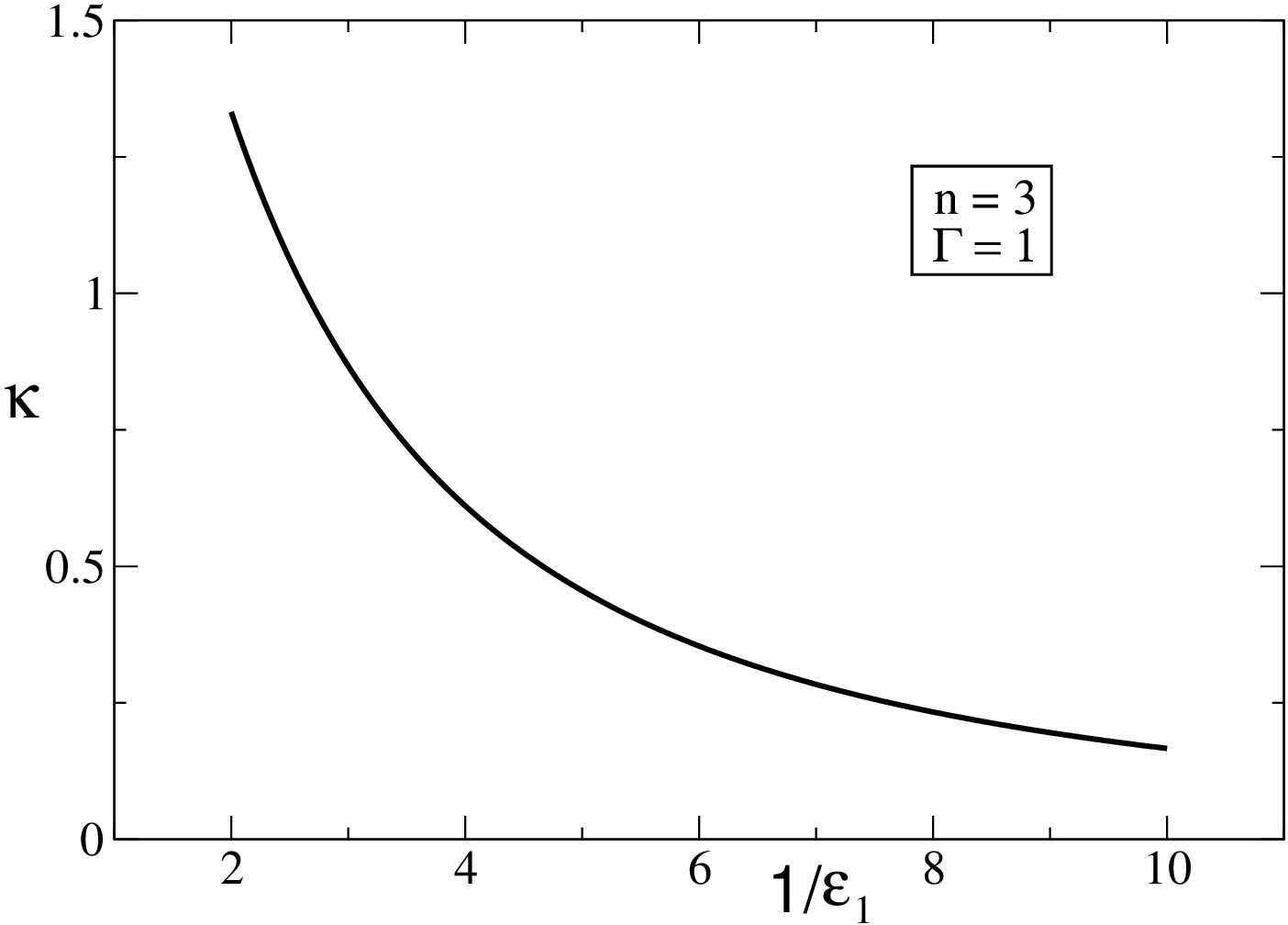}
\caption{Thermal conductivity $\kappa$, in arbitrary units at a fixed temperature $T=1$, as a function of $1/\varepsilon_1$. All energies are in units of $\Gamma_0$.}
\label{fig3}
\end{figure}

As mentioned in the introduction, an efficient thermoelectric device requires thermal conductivity $\kappa$ to be as small as possible. Presence of localized phonons due to surface disorder can suppress $\kappa$ by reducing transmission at the localized phonon frequencies.  
The linear response thermal conductivity $\kappa$ is given by the Landauer formula \cite{blencowe}
\bea
\kappa = \int_0^{\infty} \frac{d\omega}{2\pi} K(\omega); \;\;\;
K(\omega) \equiv  \omega  \mathcal{T}(\omega)\frac{\partial \eta(\omega)}{\partial T}
\label{kappa}
\eea 
where $\eta(\omega)=1/[\exp(\omega/k_BT)-1]$ is the Bose distribution function and $k_B$ is the Boltzmann constant.
We use the transmission function used in Figure \ref{fig2} with $\Gamma=1$ to evaluate $\kappa$ as a function of the inverse of the roughness power spectrum parameter, $1/\varepsilon_1$. Figure \ref{fig3} shows a significant decrease in $\kappa$ with decreasing $\varepsilon_1$. As mentioned before, either a smaller roughness correlation length $L_c$ or a larger rms surface profile $\Delta$ is expected to decrease $\kappa$; our model suggests that a single combination $1/\varepsilon_1=\Delta^2/L_c$, where $\Delta \propto (h/d)$ ($h$ is the rms height of the roughness and $d$ is the diameter of the wire), characterizes the surface disorder of the wire as far as the thermal conductivity is concerned. 

% DISCUSSION:
%=================

\section{Discussion}

Before a quantitative comparison of the thermal conductivity with experiments can be made as a function of temperature and disorder, the model needs to be improved in several different ways. First, it would require including various scattering mechanisms that are important at room temperature, and also the effects of changes in the phonon dispersion relations. Second, it would be important to include a realistic distribution of the parameters $\Omega$ and $\Gamma$ due to the random surface potential.
In addition, for the purpose of a thermoelectric device to be operated in the non-linear regime \cite{mh} which has been a major motivation for our search of a nanowire with low thermal conductivity, an evaluation of the full non-linear thermal current at finite temperature within perhaps a self-consistent Born approximation for the non-equilibrium Green functions would be necessary to understand the effects of strong phonon-surface roughness interaction. Nevertheless, the most significant results of the model are already evident at the simplest level considered here. First, the localized phonons generated by surface disorder lead to a strong frequency dependent scattering time as shown in Figure \ref{fig1}. For an arbitrary distribution of the localized phonon frequencies, these frequency dependent scatterings lead to dips in the transmission function corresponding to the localized modes, as shown in Figure \ref{fig2}. Careful experiments on such transmission function can provide important insights into the role of localized phonons in the phonon-surface roughness interactions. Finally, the importance of the power spectrum parameter $\varepsilon_1$, which can be identified with a combination of the surface roughness correlation length $L_c$ and the rms value of the roughness profile $\Delta$ is evident from the significant decrease of the thermal conductivity $\kappa$ with decreasing $\varepsilon_1$ shown in Figure \ref{fig3}.

In summary, by constructing and analyzing a simple analytically tractable model of phonon-surface roughness interaction,  we provide a theoretical understanding of why certain special type of surface disorder in Si nanowires might be more effective in suppressing phonon transport. Frequency dependent scattering off localized phonons in the model coupled to the non-trivial power spectrum of surface roughness lead to observable structure in the transmission function. This should allow experiments to check the importance of the localized phonon modes discussed in the model. We hope that the understanding developed from the model on the role of both the localized phonons and the power spectrum function of the surface disorder will help in designing an efficient thermoelectric device based on nanowires.

We acknowledge helpful discussions with S. Hershfield and D. Maslov.

\end{document}